\documentclass[aps,prc,twocolumn,superscriptaddress,nofootinbib,amsmath,amssymb,]{revtex4-2}

\usepackage{graphicx}
\usepackage{dcolumn}
\usepackage{bm}
\usepackage{float}

\usepackage[colorlinks=true, linkcolor=blue, citecolor=blue, urlcolor=blue]{hyperref}

\usepackage{lineno} 

\begin{document}

\title{Study of $p_\mathrm{T}$-differential radial flow in blast-wave model}

\author{Swati Saha}
\email{swati.saha@niser.ac.in}
\affiliation{School of Physical Sciences, National Institute of Science Education and Research, Jatni 752050, Odisha, India}
\affiliation{Homi Bhabha National Institute, Training School Complex, Anushaktinagar, Mumbai 400094, Maharashtra, India}

\author{Ranbir Singh}
\email{ranbir.singh@niser.ac.in}
\affiliation{School of Physical Sciences, National Institute of Science Education and Research, Jatni 752050, Odisha, India}
\affiliation{Homi Bhabha National Institute, Training School Complex, Anushaktinagar, Mumbai 400094, Maharashtra, India}

\author{Bedangadas Mohanty}
\email{bedanga@niser.ac.in}
\affiliation{School of Physical Sciences, National Institute of Science Education and Research, Jatni 752050, Odisha, India}
\affiliation{Homi Bhabha National Institute, Training School Complex, Anushaktinagar, Mumbai 400094, Maharashtra, India}



\begin{abstract}
The transverse momentum–differential radial flow observable $v_0(p_\mathrm{T})$, recently proposed and measured by the ATLAS and ALICE collaborations, provides a novel tool to probe radial expansion dynamics in high-energy heavy-ion collisions. In this work, we conduct a detailed study of $v_0(p_\mathrm{T})$ using a blast-wave model that incorporates hydrodynamic-like expansion and thermal emission. We introduce event-by-event fluctuations in the transverse expansion velocity and kinetic freeze-out temperature using Gaussian probability distributions. Our results show that increasing the mean expansion velocity leads to a clear mass ordering in $v_0(p_\mathrm{T})$, while fluctuations in both expansion velocity and freeze-out temperature significantly enhance the magnitude of $v_0(p_\mathrm{T})$, particularly at higher $p_\mathrm{T}$. We fit blast-wave model calculations for identified hadrons ($\pi$, K, and p) to recent ALICE data from Pb--Pb collisions at $\sqrt{s_\mathrm{NN}}$ = 5.02 TeV using a Bayesian parameter estimation framework. The extracted mean transverse expansion velocity decreases, while the kinetic freeze-out temperature increases, from central to peripheral collisions. Additionally, the freeze-out temperatures inferred from $v_0(p_\mathrm{T})$ are systematically higher than those obtained from conventional $p_\mathrm{T}$-spectra fits, likely due to the reduced sensitivity of $v_0(p_\mathrm{T})$ to resonance decay contributions.
\end{abstract}

\maketitle


\section{Introduction}
\label{sec:intro}

Heavy-ion collisions at the Large Hadron Collider (LHC) and the Relativistic Heavy Ion Collider (RHIC) create extreme conditions in which quarks and gluons become deconfined, forming a hot, dense medium known as the quark–gluon plasma (QGP)~\cite{Shuryak:1978ij, Cleymans:1985wb, ALICE:2014sbx}. As the QGP evolves, strong pressure gradients drive its collective expansion, which is well described by relativistic hydrodynamics~\cite{Ollitrault:2007du}. This expansion gives rise to two distinct types of collective motion: anisotropic flow, which reflects momentum anisotropies originating from the initial geometric asymmetries in the collision zone, and radial flow, which corresponds to the isotropic outward push of particles resulting from the system's overall pressure buildup~\cite{Ollitrault:1992bk, Voloshin:1994mz, Reisdorf:1997fx, Heinz:2013th}.

While anisotropic flow has been extensively studied through azimuthal correlations among final-state particles~\cite{STAR:2000ekf, PHENIX:2004vcz, ALICE:2010suc, ALICE:2011ab, ATLAS:2012at, CMS:2013wjq, ALICE:2016ccg}, investigations of radial flow have been comparatively limited. Recent progress has been made through studies of mean transverse momentum fluctuations~\cite{ALICE:2023tej, ALICE:2024apz, ATLAS:2024jvf, Bozek:2017elk, Giacalone:2020lbm, Samanta:2023kfk}.

Radial flow, associated with the isotropic expansion of the system, is traditionally studied via transverse momentum ($p_\mathrm{T}$) spectra, using simultaneous fits to the distributions of pions, kaons, and protons within the framework of the Boltzmann-Gibbs blast-wave model~\cite{Schnedermann:1993ws, STAR:2017sal, ALICE:2019hno}. However, this approach provides only a single value of the radial flow parameter for a given centrality class and does not capture its $p_\mathrm{T}$-dependent behavior.

The recently introduced observable $v_{0}(p_\mathrm{T})$ enables a $p_\mathrm{T}$-differential study of radial flow by capturing long-range transverse momentum correlations while effectively suppressing short-range nonflow effects through the use of a pseudorapidity gap~\cite{Schenke:2020uqq, Parida:2024ckk}. Defined via a normalized covariance between the event-wise mean $p_\mathrm{T}$ and the particle yield in $p_\mathrm{T}$ bins, $v_{0}(p_\mathrm{T})$ offers a novel probe of collective dynamics in the QGP. Measurements by both the ALICE and ATLAS collaborations for inclusive charged particles in Pb--Pb collisions at $\sqrt{s_\mathrm{NN}} = 5.02$~TeV have demonstrated its sensitivity to key medium properties such as bulk viscosity and the QCD equation of state~\cite{ALICE:2025iud, ATLAS:2025ztg}. Recent studies employing full (3+1)D hydrodynamic simulations using the MUSIC framework have shown that $v_{0}(p_\mathrm{T})$ for identified particles exhibits a clear mass ordering at low $p_\mathrm{T}$. This behavior, attributed to collective expansion dynamics, is analogous to the well-known mass ordering observed in the elliptic flow coefficient $v_{2}(p_\mathrm{T})$~\cite{Schenke:2020uqq, Parida:2024ckk}. In such models, the magnitude and shape of $v_{0}(p_\mathrm{T})$ are found to be sensitive to initial-state fluctuations, transport coefficients (such as bulk viscosity), and freeze-out conditions. ALICE measurements confirm the mass-dependent behavior at low $p_\mathrm{T}$ and also reveal baryon–meson separation at intermediate $p_\mathrm{T}$~\cite{ALICE:2025iud}. These observations establish $v_{0}(p_\mathrm{T})$ as a sensitive tool for studying collective expansion, providing complementary insights beyond those available from conventional flow observables.

In this work, we analyze $v_{0}(p_\mathrm{T})$ for pions, kaons, and protons within the theoretical framework of the Boltzmann-Gibbs blast-wave model~\cite{Schnedermann:1993ws} for Pb--Pb collisions at $\sqrt{s_\mathrm{NN}} = 5.02$~TeV. The blast-wave model provides a simplified, hydrodynamically motivated description of the kinetic freeze-out stage, in which a locally thermalized medium undergoes collective radial expansion. In this framework, the transverse momentum spectra of final-state particles are governed primarily by two parameters: the radial flow velocity, which reflects the collective expansion strength, and the kinetic freeze-out temperature, which characterizes the thermal conditions at decoupling. This study establishes a connection between the traditional extraction of radial flow and freeze-out temperature via hadron $p_\mathrm{T}$ spectra and the corresponding results derived from $v_{0}(p_\mathrm{T})$. Additionally, we aim to disentangle the contributions of thermal motion and collective flow to the behavior of $v_{0}(p_\mathrm{T})$.

The paper is organized as follows: Section~\ref{sec:obs} defines the $v_{0}(p_\mathrm{T})$ observable and outlines the analysis methodology; Section~\ref{sec:res} presents the results of the blast-wave model calculations and compares them with ALICE measurements; and Section~\ref{sec:summ} summarizes the main findings and their implications.

\begin{figure*}
\centering
\includegraphics[width=\textwidth]{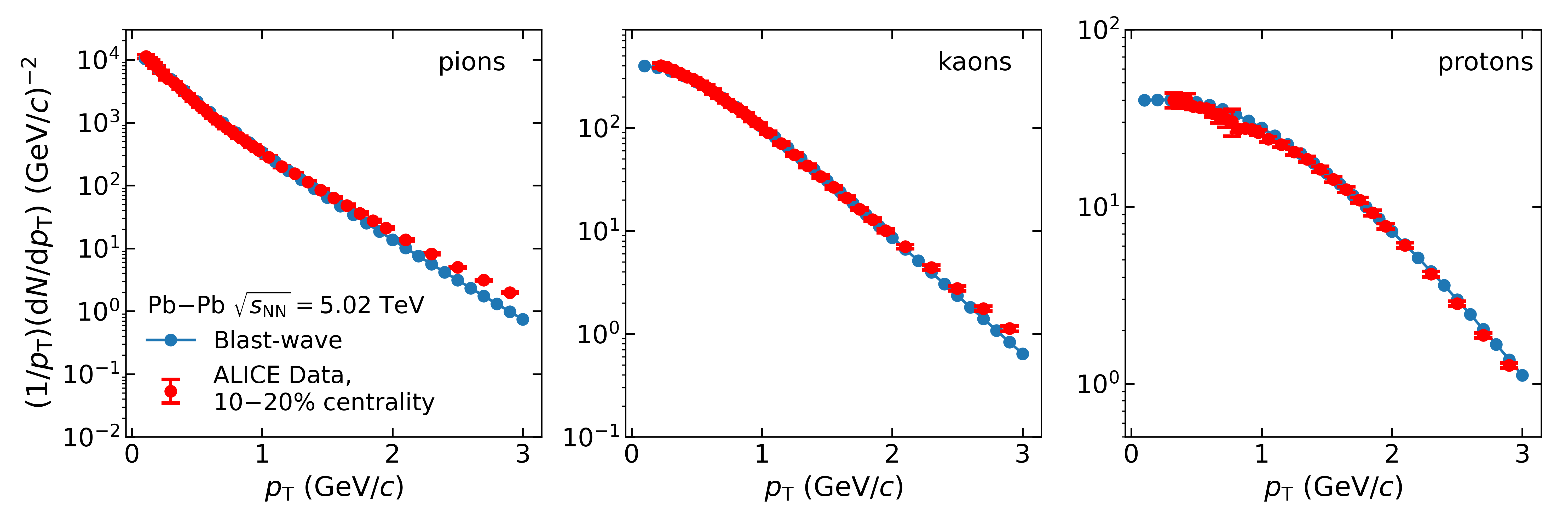}
\caption{Transverse momentum spectra of pions (left), kaons (middle), and protons (right) measured in Pb$-$Pb collisions at $\sqrt{s_\mathrm{NN}} = 5.02$ TeV for centrality class 10$-$20\% is compared with blast-wave model generated results using the parameters $\beta_s = 0.897$, $T_\text{kin} = 0.094$ GeV, and $n=0.739$ from Ref.~\cite{ALICE:2019hno}.}
\label{fig:spectra}
\end{figure*}

\section{Observable and methodology}
\label{sec:obs}
We follow the definition of $v_{0}(p_\mathrm{T})$ from Refs.~\cite{Schenke:2020uqq, Parida:2024ckk}, which characterizes event-by-event fluctuations in the transverse momentum spectra:

\begin{equation}
v_{0}(p_\mathrm{T}) = \frac{\langle \delta f(p_\mathrm{T}) \, \delta [p_\mathrm{T}] \rangle}{\langle f(p_\mathrm{T}) \rangle \, \sigma([p_\mathrm{T}])}.
\end{equation}

Here, $[p_\mathrm{T}]$ denotes the event-by-event mean transverse momentum of charged particles, and $f(p_\mathrm{T})$ is the normalized transverse momentum distribution, related to the event-by-event $p_\mathrm{T}$ spectra by

\begin{equation}
f(p_\mathrm{T}) = \frac{N(p_\mathrm{T})}{\int N(p_\mathrm{T}) \, \mathrm{d}p_\mathrm{T}}.
\end{equation}

The quantities $\delta f(p_\mathrm{T})$ and $\delta [p_\mathrm{T}]$, which quantify the fluctuations of $f(p_\mathrm{T})$ and $[p_\mathrm{T}]$ about their event averages, are given by

\begin{equation}
\delta f(p_\mathrm{T}) = f(p_\mathrm{T}) - \langle f(p_\mathrm{T}) \rangle, \qquad \delta [p_\mathrm{T}] = [p_\mathrm{T}] - \langle [p_\mathrm{T}] \rangle.
\end{equation}

In the Boltzmann-Gibbs blast-wave model, particle production is described by

\begin{equation}
\label{eq:blast}
E \frac{\mathrm{d}^3N}{\mathrm{d} p^3} \propto \int_0^R m_\mathrm{T} I_0\left(\frac{p_\mathrm{T} \sinh(\rho)}{T_{\text{kin}}}\right) 
K_1\left(\frac{m_\mathrm{T} \cosh(\rho)}{T_{\text{kin}}}\right) r\,\mathrm{d}r,
\end{equation}

where $m_\mathrm{T}$ is the transverse mass, defined as $m_\mathrm{T} = \sqrt{m^2 + p_\mathrm{T}^2}$, $T_{\text{kin}}$ is the temperature at kinetic freeze-out, $I_0$ and $K_1$ are modified Bessel functions, and $r$ is the radial distance in the transverse plane. Equation~\ref{eq:blast} is used to generate event-by-event $p_\mathrm{T}$ spectra for estimating $v_{0}(p_\mathrm{T})$. The radial expansion velocity profile $\rho$ is parametrized as

\begin{equation}
\rho = \tanh^{-1} \beta_\mathrm{T}(r) = \tanh^{-1} \left[ \left( \frac{r}{R} \right)^n \beta_s \right],
\end{equation}

where $R$ is the fireball radius, $\beta_\mathrm{T}(r)$ is the radial velocity, and $\beta_s$ is the transverse velocity at the fireball surface. The mean radial expansion velocity $\langle\beta_\mathrm{T}\rangle$ is related to $\beta_s$ by

\begin{equation}
\label{eq:betaS}
\langle \beta_\mathrm{T} \rangle = \frac{\int_0^R \beta_\mathrm{T}(r)\,r\,\mathrm{d}r}{\int_0^R r\,\mathrm{d}r} \implies \beta_s = \langle \beta_\mathrm{T} \rangle \left(\frac{n+2}{2}\right).
\end{equation}

Thus, for a given value of $\langle\beta_\mathrm{T}\rangle$, the corresponding $\beta_s$ can be determined using Eq.~\ref{eq:betaS}. The Boltzmann-Gibbs blast-wave function (Eq.~\ref{eq:blast}) is typically used to simultaneously fit the $p_\mathrm{T}$ spectra of charged pions, kaons, and protons using a common parameter set, while allowing for separate normalization factors and masses. 

The parameter set reported in Ref.~\cite{ALICE:2019hno} is used to generate event-by-event blast-wave spectra, and the averaged spectrum is compared with experimental data in Fig.~\ref{fig:spectra}. The spectra are obtained for the 10--20\% centrality class using the parameter values $\langle\beta_\mathrm{T}\rangle = 0.655$ (yielding $\beta_s = 0.897$), $T_\text{kin} = 0.094$~GeV, and $n = 0.739$. The normalization factors are adjusted to match the experimental spectra, ensuring a reliable estimate of the mean $p_\mathrm{T}$ for inclusive charged particles.

These generated blast-wave spectra, tuned to reproduce the measured $p_\mathrm{T}$ distributions of pions, kaons, and protons, provide a realistic baseline to investigate how event-by-event fluctuations in the radial flow velocity and kinetic freeze-out temperature influence the behavior of $v_{0}(p_\mathrm{T})$. While the mean values of these parameters determine the shape of the average spectra, their fluctuations manifest in $v_{0}(p_\mathrm{T})$.

To systematically quantify these effects, we introduce Gaussian fluctuations in $\beta_{s}$ and $T_{\text{kin}}$ and consider four scenarios in which either the mean or the width of each parameter is varied independently:
(i) varying the mean of $\beta_{s}$ while keeping its fluctuation width and $T_{\text{kin}}$ fixed,
(ii) varying the fluctuation width of $\beta_{s}$ while keeping its mean and $T_{\text{kin}}$ fixed,
(iii) varying the mean of $T_{\text{kin}}$ while $\beta_{s}$ is held constant, and
(iv) varying the fluctuation width of $T_{\text{kin}}$ while its mean and $\beta_{s}$ are kept fixed.

These controlled variations allow us to probe the respective roles of thermal motion and collective flow fluctuations in shaping the structure of $v_{0}(p_\mathrm{T})$. In the following section, we present the results of our analysis, emphasizing the sensitivity of $v_{0}(p_\mathrm{T})$ to variations in blast-wave parameters.

\section{Results and discussion}
\label{sec:res}

\begin{figure*}
\centering
\includegraphics[width=\textwidth]{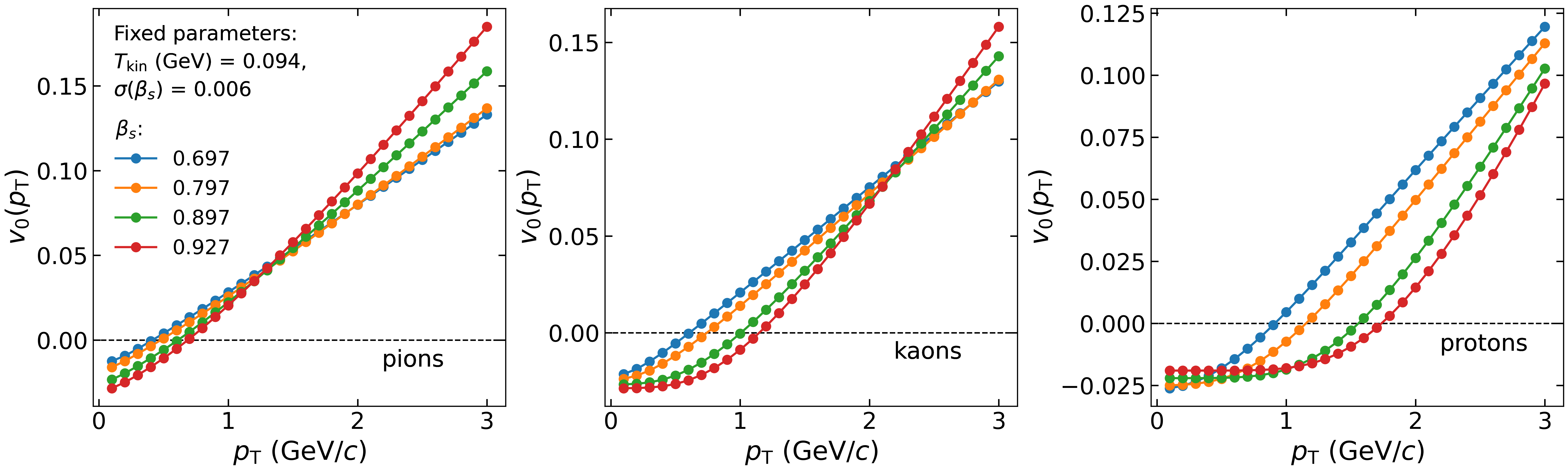}
\caption{$v_{0}(p_\mathrm{T})$ for pions (left), kaons (middle), and protons (right) from the blast-wave model incorporating Gaussian fluctuations in the transverse expansion velocity at the surface. The kinetic freeze-out temperature ($T_\mathrm{kin}$) and the fluctuation width of the surface velocity $\sigma(\beta_s)$ are fixed at 0.094 GeV and 0.006, respectively, while the mean surface velocity $\beta_s$ is varied in the range [0.697, 0.927].}
\label{fig:varyBetaSmean}
\end{figure*}

The effect of varying the surface transverse expansion velocity $\beta_s$ on the evolution of $v_{0}(p_\mathrm{T})$ for individual particle species—pions (left), kaons (middle), and protons (right)—is shown in Fig.~\ref{fig:varyBetaSmean}. The results correspond to four different $\beta_s$ values ranging from 0.697 to 0.927, with a fixed kinetic freeze-out temperature $T_\mathrm{kin} = 0.094$~GeV and a fluctuation width of $\sigma(\beta_{s}) = 0.006$.

For all particle species, the general trend is that $v_{0}(p_\mathrm{T})$ transitions from negative values at low $p_\mathrm{T}$ to positive values beyond a threshold transverse momentum, $p_\mathrm{T,sep}$. This threshold increases with $\beta_s$, indicating that higher expansion velocities shift the onset of positive $v_{0}(p_\mathrm{T})$ to larger $p_\mathrm{T}$. The increase in $p_\mathrm{T,sep}$ with $\beta_s$ is most significant for protons, followed by kaons and then pions, indicating that stronger radial flow enhances the mass-dependent separation among particle species.

The left panel presents results for pions, where $v_{0}(p_\mathrm{T})$ increases almost linearly for all $\beta_s$. For $p_\mathrm{T} < 1.5$~GeV/$c$, higher $\beta_s$ values lead to a reduction in $v_{0}(p_\mathrm{T})$, whereas for $p_\mathrm{T} > 1.5$~GeV/$c$, increasing $\beta_s$ results in an enhancement. Similar results for kaons, shown in the middle panel, reveal a more nonlinear dependence on $p_\mathrm{T}$ at higher $\beta_s$ values compared to pions. For $p_\mathrm{T} < 2.5$~GeV/$c$, increasing $\beta_s$ reduces $v_{0}(p_\mathrm{T})$, following a trend similar to that of pions but with greater non-linearity. The transition point where $v_{0}(p_\mathrm{T})$ begins to increase with $\beta_s$ occurs at a higher $p_\mathrm{T}$ than for pions. Beyond this point, $v_{0}(p_\mathrm{T})$ rises more steeply with $p_\mathrm{T}$ as $\beta_s$ increases, although the overall change in magnitude remains modest.

The right panel displays results for protons, which exhibit the strongest nonlinear response to changes in $\beta_s$ among the three species. At the lowest $\beta_s$, $v_{0}(p_\mathrm{T})$ follows a linear trend with $p_\mathrm{T}$, but as $\beta_s$ increases, the dependence becomes increasingly nonlinear—more so than for kaons. For $p_\mathrm{T} < 0.5$~GeV/$c$, the effect of $\beta_s$ is minimal, whereas for $p_\mathrm{T} > 1$~GeV/$c$, larger $\beta_s$ leads to a stronger suppression of $v_{0}(p_\mathrm{T})$, consistent with the trends observed for pions and kaons.

\begin{figure*}
\centering
\includegraphics[width=\textwidth]{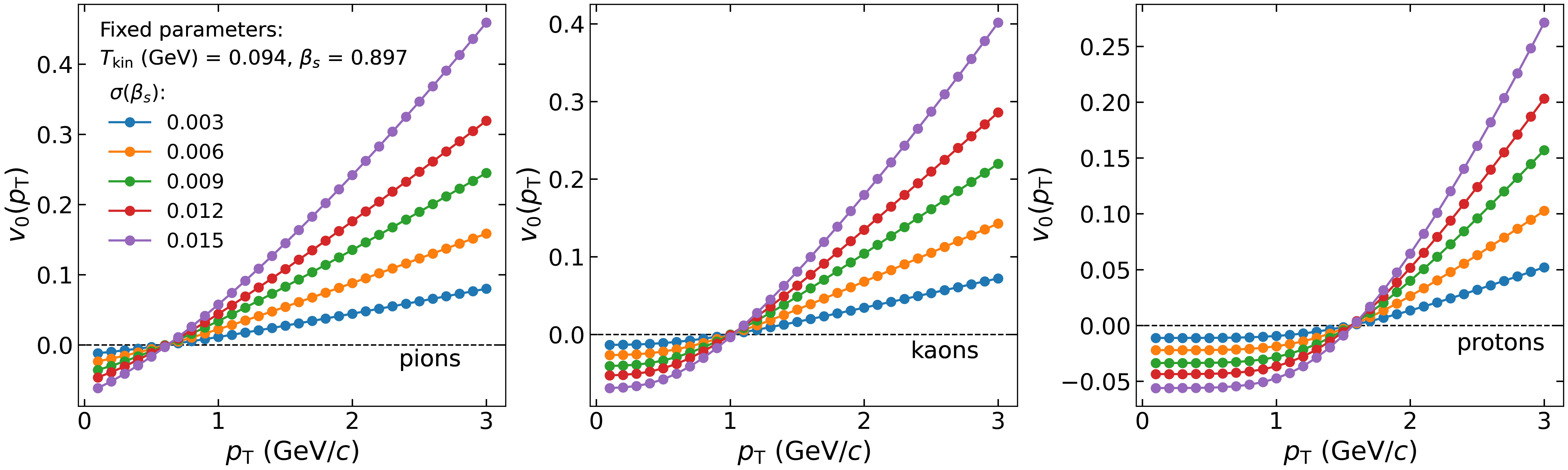}
\caption{$v_{0}(p_\mathrm{T})$ of pions (left), kaons (middle), and protons (right) in blast-wave model for Gaussian-fluctuations of transverse expansion velocity at the surface. The kinetic freeze-out temperature ($T_\mathrm{kin}$) and mean value of fluctuations $\beta_s$ are kept fixed at 0.094 GeV and 0.897, respectively, while the width $\sigma(\beta_s)$ is varied in the range [0.003, 0.015]. }
\label{fig:varyBetaSsigma}
\end{figure*}

\begin{figure*}
\centering
\includegraphics[width=\textwidth]{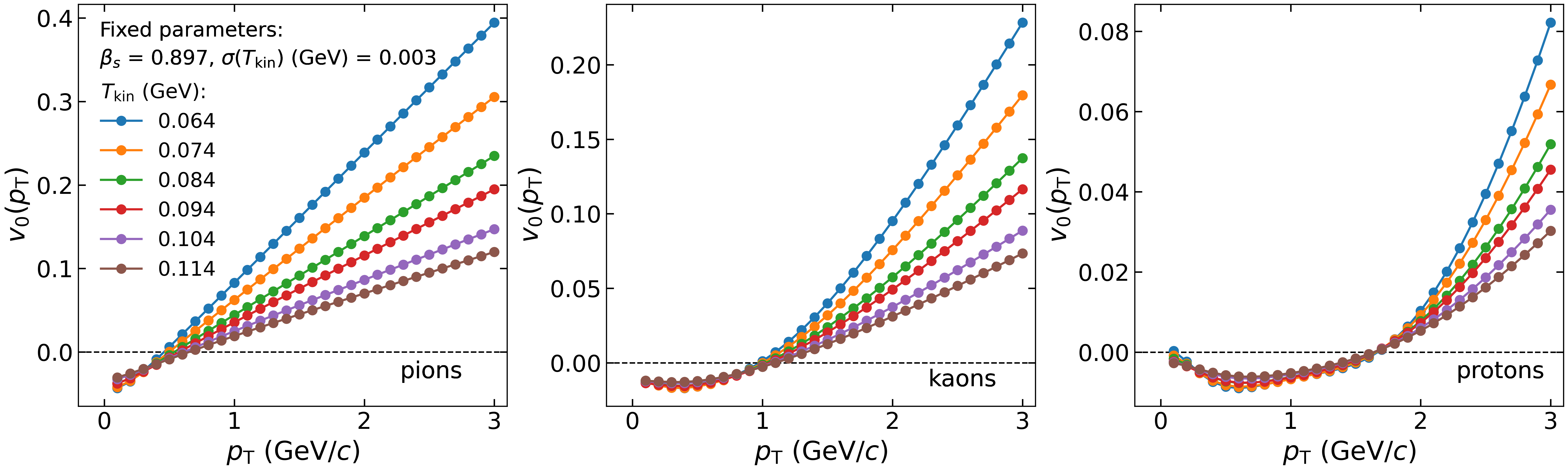}
\caption{$v_{0}(p_\mathrm{T})$ of pions (left), kaons (middle), and protons (right) in blast-wave model for Gaussian-fluctuations of kinetic freeze-out temperature. The transverse expansion velocity at the
surface ($\beta_s$) and width of fluctuations $\sigma(T_\mathrm{kin})$ are kept fixed at 0.897 and 0.003 GeV, respectively, while the mean value $T_\mathrm{kin}$ is varied in the range [0.064, 0.114] GeV.}
\label{fig:varyTkinMean}
\end{figure*}

Figure~\ref{fig:varyBetaSsigma} illustrates the behavior of $v_{0}(p_\mathrm{T})$ for pions (left), kaons (middle), and protons (right), incorporating Gaussian fluctuations in $\beta_s$. The kinetic freeze-out temperature $T_\mathrm{kin}$ and the mean value of $\beta_s$ are held constant, while the width of the fluctuations, $\sigma(\beta_s)$, is systematically varied from 0.003 to 0.015. The results indicate a clear dependence of $v_{0}(p_\mathrm{T})$ on $\sigma(\beta_s)$ for all three particle species.

Increasing fluctuations in $\beta_s$ lead to an overall increase in the magnitude of $v_{0}(p_\mathrm{T})$ both below and above $p_\mathrm{T,sep}$. This effect is more significant for $p_\mathrm{T} > p_\mathrm{T,sep}$ than for $p_\mathrm{T} < p_\mathrm{T,sep}$ and becomes more pronounced with increasing $p_\mathrm{T}$. Despite the variation in $\sigma(\beta_s)$, the threshold $p_\mathrm{T,sep}$—which marks the transition between low- and high-$p_\mathrm{T}$ behavior—remains invariant across all three particle species. The slope of $v_{0}(p_\mathrm{T})$ at high $p_\mathrm{T}$ is steepest for protons, followed by kaons and then pions, suggesting that the impact of $\beta_s$ fluctuations is stronger for heavier particles. As protons are more massive than kaons and pions, they exhibit greater sensitivity to variations in $\beta_s$.

The effect of Gaussian fluctuations in $T_\mathrm{kin}$ on the magnitude of $v_{0}(p_\mathrm{T})$ is illustrated in Fig.~\ref{fig:varyTkinMean}. The values of $\beta_s$ and the fluctuation width of $T_\mathrm{kin}$, $\sigma(T_\mathrm{kin})$, are fixed at 0.897 and 0.003~GeV, respectively, while the mean value of $T_\mathrm{kin}$ is systematically varied. The results for pions (left), kaons (middle), and protons (right) are shown for $T_\mathrm{kin}$ values ranging from 0.064~GeV to 0.114~GeV.

No significant change in $v_{0}(p_\mathrm{T})$ is observed below the corresponding $p_\mathrm{T,sep}$ for each particle species. However, for $p_\mathrm{T} > p_\mathrm{T,sep}$, $v_{0}(p_\mathrm{T})$ decreases with increasing $T_\mathrm{kin}$ across all species, with the effect becoming more pronounced at higher $p_\mathrm{T}$. This trend suggests that a higher $T_\mathrm{kin}$—corresponding to increased thermal motion—tends to dilute fluctuations in the $p_\mathrm{T}$ spectra, thereby reducing $v_{0}(p_\mathrm{T})$.

Furthermore, the mass hierarchy and separation among particle species in $v_{0}(p_\mathrm{T})$ remain largely unaffected by variations in $T_\mathrm{kin}$, in contrast to the stronger sensitivity observed with changes in $\beta_s$, which more directly influence these effects.

\begin{figure*}
\centering
\includegraphics[width=\textwidth]{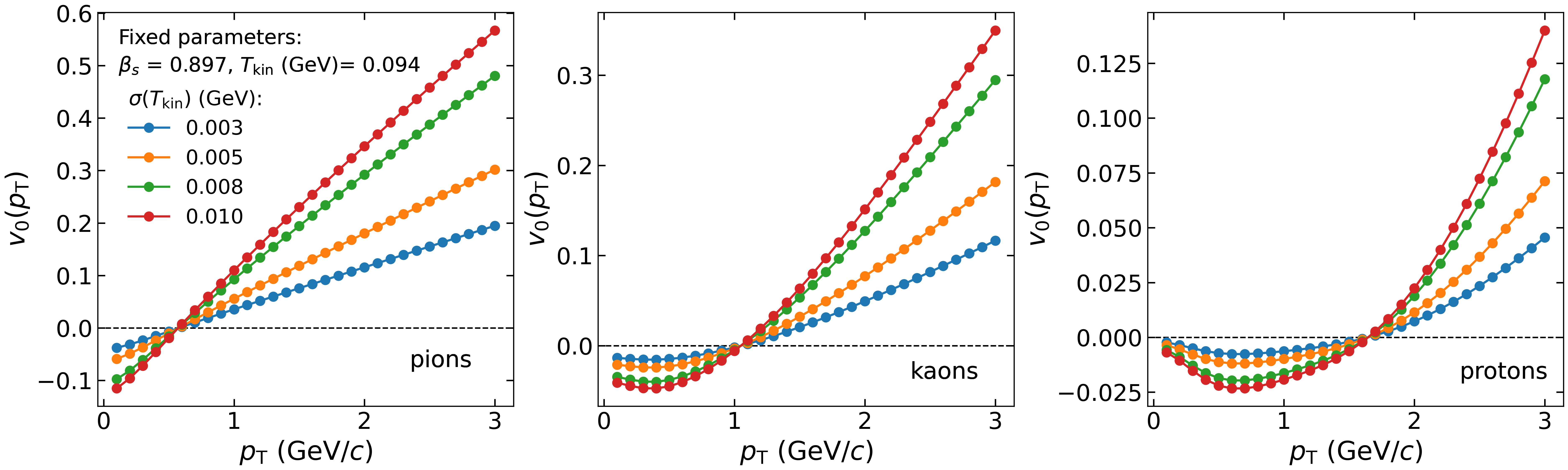}
\caption{$v_{0}(p_\mathrm{T})$ of pions (left), kaons (middle), and protons (right) in blast-wave model for Gaussian-fluctuations of kinetic freeze-out temperature. The transverse expansion velocity at the
surface ($\beta_s$) and mean value of fluctuations $T_\mathrm{kin}$ are kept fixed at 0.897 and 0.094 GeV, respectively, while the width $\sigma(T_\mathrm{kin})$ is varied in the range [0.003, 0.010] GeV.}
\label{fig:varyTkinSigma}
\end{figure*}

\begin{table*}
\centering
\caption{Summary of the key findings from the blast-wave model investigation of $v_{0}(p_\mathrm{T})$ by the variation of mean values for transverse expansion velocity ($\beta_s$), kinetic freeze-out temperature ($T_\mathrm{kin}$), and their fluctuations ($\sigma(\beta_s)$ and $\sigma(T_\mathrm{kin}$)).}
\label{tab:summary}
\begin{tabular}{l l}
\hline\hline
Parameter increased & \multicolumn{1}{c}{Observed Trends} \\
\hline\hline
$\beta_s$ (mean) & - Increases the difference in $v_{0}(p_\mathrm{T})$ between particle species\\
& - $v_{0}(p_\mathrm{T})$ increases at high $p_\mathrm{T}$ but decreases at low $p_\mathrm{T}$\\
& - $p_\mathrm{T,sep}$ shifts to higher values, effect more pronounced for heavier particles\\
\hline
$\sigma(\beta_s)$ & - Enhances $v_{0}(p_\mathrm{T})$ across the whole $p_\mathrm{T}$ range\\
& - No change in the position of $p_\mathrm{T,sep}$\\
\hline
$T_\mathrm{kin}$ (mean) & - No change in the difference in $v_{0}(p_\mathrm{T})$ between particle species\\
& - $v_{0}(p_\mathrm{T})$ decreases above $p_\mathrm{T,sep}$, no change in the magnitude below $p_\mathrm{T,sep}$\\
& - No change in the position of $p_\mathrm{T,sep}$\\
\hline
$\sigma(T_\mathrm{kin}$) & - Increases $v_{0}(p_\mathrm{T})$, particularly at high $p_\mathrm{T}$\\
& - No change in the position of $p_\mathrm{T,sep}$\\
\hline
\hline
\end{tabular}
\end{table*}

Figure~\ref{fig:varyTkinSigma} illustrates the dependence of $v_{0}(p_\mathrm{T})$ on $p_\mathrm{T}$ for pions (left), kaons (middle), and protons (right). The primary focus of this figure is to examine the effect of varying the width of Gaussian fluctuations in $T_\mathrm{kin}$. The fluctuation width, $\sigma(T_\mathrm{kin})$, is varied from 0.003~GeV to 0.01~GeV, while $\beta_s$ is fixed at 0.897 and the mean value of $T_\mathrm{kin}$ is held constant at 0.094~GeV. 

The results reveal a strong dependence of $v_{0}(p_\mathrm{T})$ on $\sigma(T_\mathrm{kin})$ across all three particle species, resembling the trends observed for $\sigma(\beta_s)$ in Fig.~\ref{fig:varyBetaSsigma}. As fluctuations in $T_\mathrm{kin}$ increase, the magnitude of $v_{0}(p_\mathrm{T})$ increases both below and above $p_\mathrm{T,sep}$. This effect is more pronounced at higher $p_\mathrm{T}$, particularly for $p_\mathrm{T} > p_\mathrm{T,sep}$, where the rise in $v_{0}(p_\mathrm{T})$ becomes steeper. Notably, the slope of $v_{0}(p_\mathrm{T})$ at high $p_\mathrm{T}$ is steepest for protons, followed by kaons and pions, indicating a stronger sensitivity of heavier particles to fluctuations in $T_\mathrm{kin}$.

Table~\ref{tab:summary} summarizes how collective expansion and thermal motion shape the structure of $v_{0}(p_\mathrm{T})$ within the blast-wave model by varying $\beta_s$, $T_\mathrm{kin}$, and their fluctuations. Although the blast-wave framework does not incorporate initial-state geometry or the full dynamical evolution of the collision, it captures the essential physics underlying mass ordering and the general behavior of $v_{0}(p_\mathrm{T})$. In particular, the model demonstrates how radial flow and its event-by-event fluctuations, along with variations in freeze-out temperature, govern the $p_\mathrm{T}$-differential behavior of $v_{0}(p_\mathrm{T})$ observed across different particle species. 

The qualitative agreement between these results and those from more complex hydrodynamic simulations suggests that the blast-wave approach serves as a simplified yet effective tool for interpreting $v_{0}(p_\mathrm{T})$. Incorporating resonance decay contributions and initial-state geometry effects will be important directions for future work, enabling a more quantitative comparison with full hydrodynamic models.

\begin{figure*}
\centering
\includegraphics[width=0.8\textwidth]{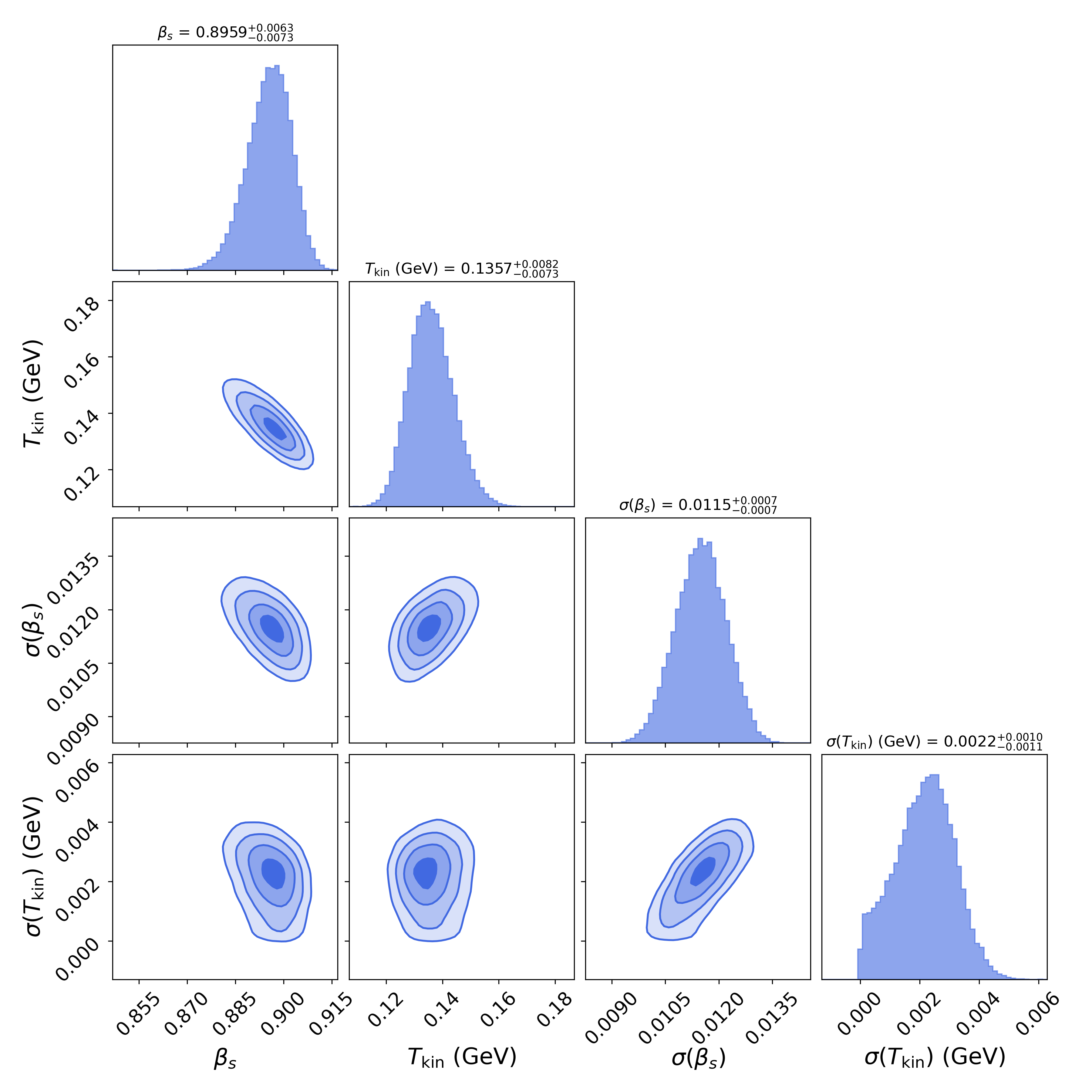}
\caption{The posterior distributions and parameter correlations for $\beta_s$, $T_\mathrm{kin}$, $\sigma(\beta_s)$, and $\sigma(T_\mathrm{kin})$ using $v_{0}(p_\mathrm{T})$ measurements of pions, kaons, and protons for centrality classes 10$-$20\% \cite{ALICE:2025iud}. The diagonal panels display the marginalized probability distributions of individual parameters, while the off-diagonal panels show the 2D joint distributions with contour levels indicating confidence regions. The plot suggests correlations between some parameters, highlighting the uncertainties in their estimations.}
\label{fig:post}
\end{figure*}

For the extraction of radial flow parameters, the blast-wave predictions of $v_{0}(p_\mathrm{T})$ for pions, kaons, and protons are compared to measurements reported by ALICE for the centrality classes 10--20\%, 30--40\%, and 60--70\%. The parameter inference is performed using Bayesian parameter estimation, a rigorous probabilistic framework that combines prior knowledge of the parameters with experimental data to update the probability distributions describing their values. This approach yields a posterior probability distribution for each model parameter via Bayes' theorem.

Markov Chain Monte Carlo (MCMC) methods are employed to efficiently sample the posterior distribution, enabling robust uncertainty quantification and capturing correlations between parameters. The posterior distributions reflect both the information content of the data and the inherent uncertainties, as shown in Fig.~\ref{fig:post}. Uniform (uninformative) priors are applied over physically motivated ranges for all model parameters, as listed in Table~\ref{tab: priors}. These bounds are chosen based on typical freeze-out conditions inferred from previous heavy-ion collision studies at LHC energies. A modest extension of these prior ranges (e.g., by $\pm$10–20\%) was tested and found not to significantly alter the posterior distributions, confirming that the results are primarily data-driven.

A single set of model parameters is used to jointly describe the $v_{0}(p_\mathrm{T})$ distributions of all three particle species within each centrality class. In the fitting procedure, the chi-squared ($\chi^2$) statistic is computed as the deviation between the blast-wave model predictions and the experimental data at the same $p_\mathrm{T}$ intervals used for fitting the $p_\mathrm{T}$ spectra in Ref.~\cite{ALICE:2019hno}. The $p_\mathrm{T}$ intervals are 0.5--1~GeV/$c$, 0.2--1.5~GeV/$c$, and 0.3--3~GeV/$c$ for charged pions, kaons, and protons, respectively.

\begin{table}
\centering
\caption{Uniform prior ranges used for the blast-wave model parameters in the Bayesian analysis of $v_0(p_\mathrm{T})$ for pions, kaons, and protons.}
\label{tab: priors}
\begin{tabular}{l l}
\hline\hline
Model parameter & Prior ranges \\
\hline
$\beta_{s}$ & [0.60, 0.92]\\
$T_\mathrm{kin}$ (GeV) & [0.08, 0.35]\\
$\sigma(\beta_{s})$ &  [0.001, 0.06]\\
$\sigma(T_\mathrm{kin})$ (GeV) & [0.0, 0.05]\\
\hline\hline
\end{tabular}
\end{table}


\begin{figure*}
  \centering
  \includegraphics[width=\textwidth]{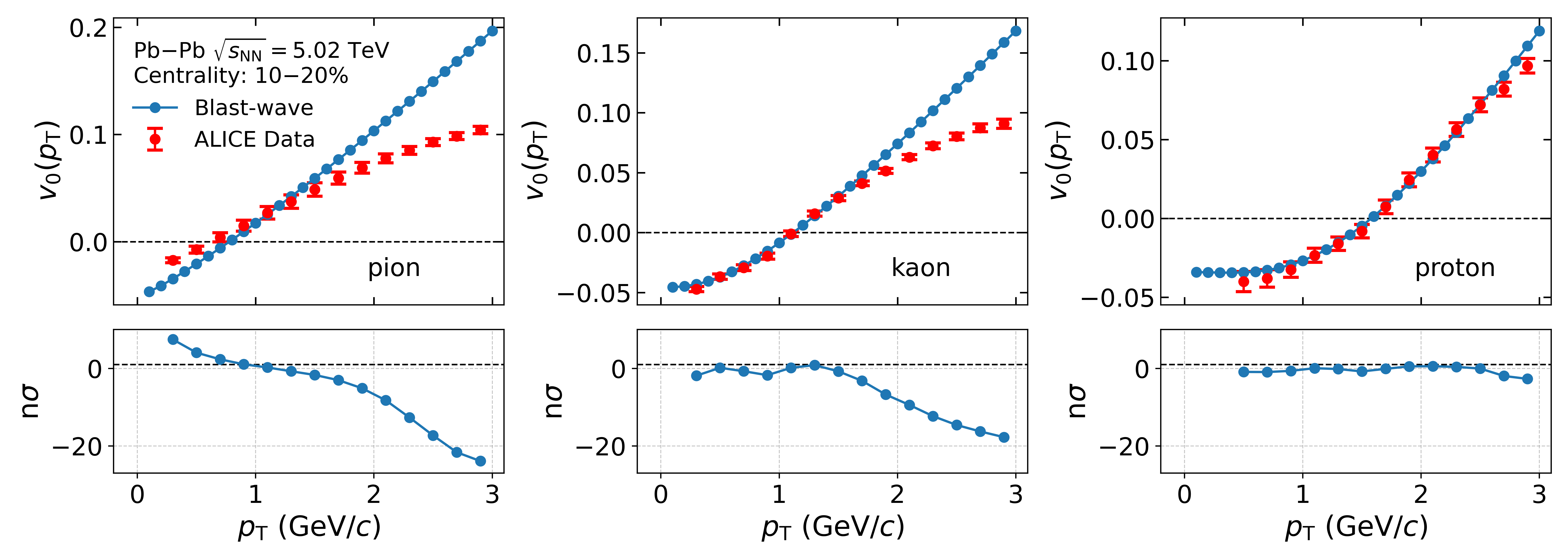}
  \includegraphics[width=\textwidth]{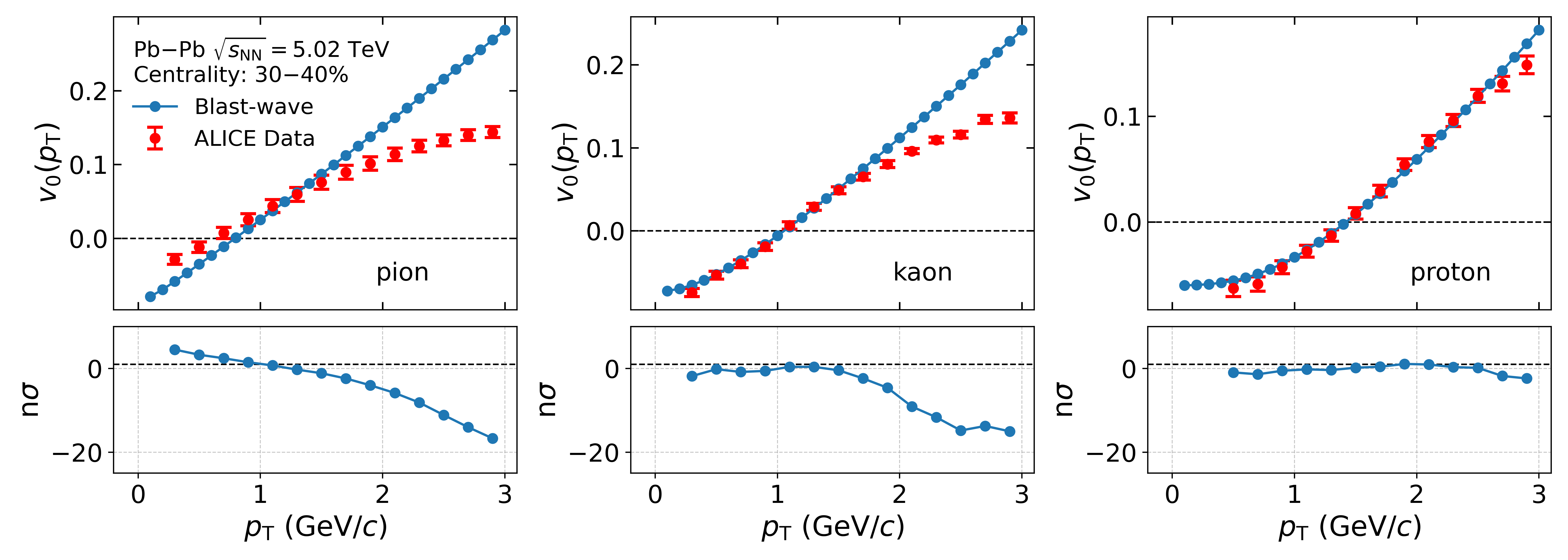}
  \includegraphics[width=\textwidth]{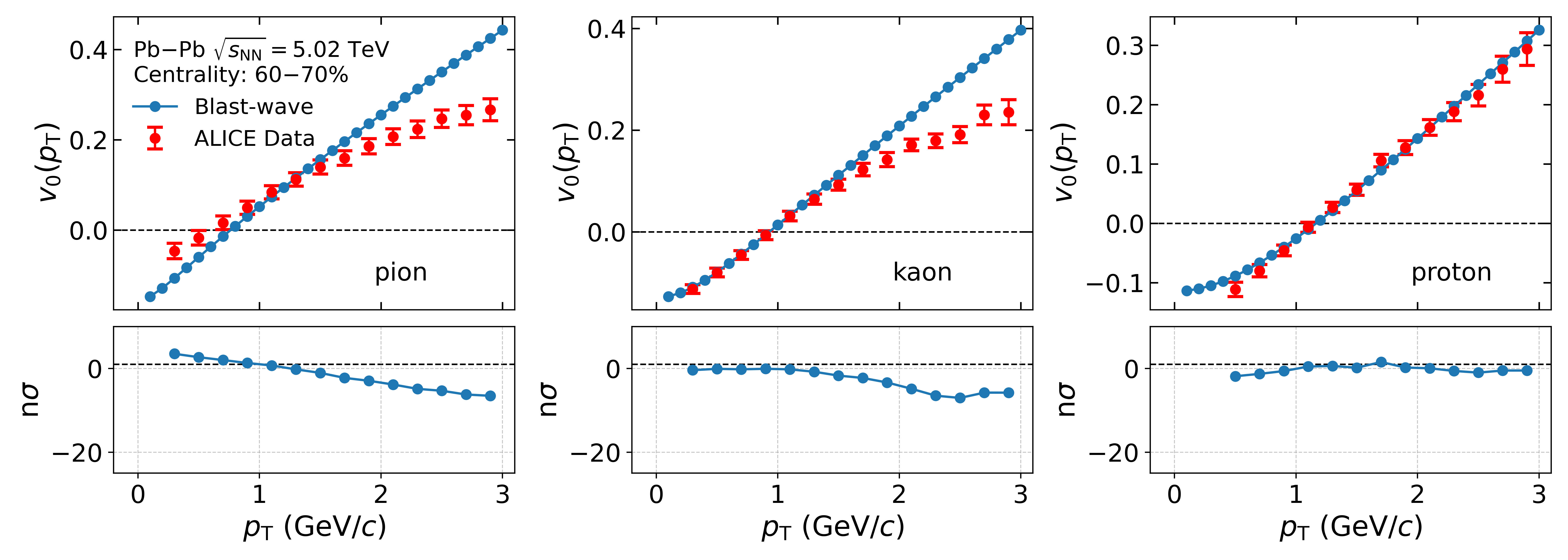}
\caption{$v_{0}(p_\mathrm{T})$ of pions (left), kaons (middle), and protons (right) in Pb$-$Pb collisions at $\sqrt{s_\mathrm{NN}} = 5.02$ TeV for centrality classes 10$-$20\%, 30$-$40\%, and 60$-$70\%. The measurements of ALICE \cite{ALICE:2025iud} (red marker) are compared to blast-wave predictions obtained with the best-fit parameters. The bottom panels show the residual, n$\sigma$ = $\mathrm{(data-model)/error_{data}}$.}
\label{fig:v0pT_data_compare}
\end{figure*}

For the 10--20\% centrality interval, the Bayesian analysis yields the following median parameter estimates with their corresponding 68\% credible intervals: $\beta_{s} = 0.8959^{+0.0063}_{-0.0073}$, $T_\mathrm{kin} = 0.1357^{+0.0082}_{-0.0073}$ GeV, and their event-by-event fluctuations are found to be $\sigma(\beta_{s}) = 0.0115^{+0.0007}_{-0.0007}$, and $\sigma(T_\mathrm{kin})=0.0022^{+0.0010}_{-0.0011}$ GeV. The model predictions of $v_{0}(p_\mathrm{T})$ computed with these best-fit parameters are shown alongside the ALICE experimental data in Fig.~\ref{fig:v0pT_data_compare} for three different centrality intervals. The smaller bottom panels in Fig.~\ref{fig:v0pT_data_compare} display the residuals normalized by the total experimental uncertainty, i.e., (Data-Model)/$\sigma_\mathrm{Data}$. A value of zero indicates perfect agreement, while deviations within 2$\sigma$ suggest consistency between data and model. This comparison in Fig.~\ref{fig:v0pT_data_compare} demonstrates that the blast-wave model with the extracted parameters successfully captures the key features of the data. A very good agreement is noted for protons, which is followed by kaons and pions.

\renewcommand{\arraystretch}{1.5}
\begin{table*}
\centering
\caption{Extracted blast-wave parameters and their event-by-event fluctuation width for different centrality intervals.}
\label{tab:parameters}
\begin{tabular}{c c c c c}
\hline\hline
Centrality & $\beta_{s}$ & $T_{\mathrm{kin}}$ (GeV) & $\sigma(\beta_{s})$ & $\sigma(T_{\mathrm{kin}})$ (GeV) \\
\hline
10--20\% & $0.8959^{+0.0063}_{-0.0073}$ & $0.1357^{+0.0082}_{-0.0073}$ & $0.0115^{+0.0007}_{-0.0007}$ & $0.0022^{+0.0010}_{-0.0011}$ \\ 
30--40\% & $0.8420^{+0.0125}_{-0.0142}$ & $0.1719^{+0.0117}_{-0.0107}$ & $0.0212^{+0.0011}_{-0.0011}$ & $0.0065^{+0.0019}_{-0.0021}$ \\
60--70\% & $0.7005^{+0.0379}_{-0.0424}$ & $0.2346^{+0.0185}_{-0.0183}$ & $0.0455^{+0.0046}_{-0.0046}$ & $0.0201^{+0.0030}_{-0.0032}$ \\
\hline
\hline
\end{tabular}
\end{table*}
\renewcommand{\arraystretch}{1.}

A detailed summary of the inferred parameters and their uncertainties for all three centrality intervals studied is compiled in Table~\ref{tab:parameters}. There is a clear decreasing trend in $\beta_s$ from central to peripheral collisions. For the 10--20\% centrality class, $\beta_s$ is approximately 0.896, indicating strong collective expansion. This value systematically decreases to about 0.842 in the 30--40\% class and further to 0.700 in the 60--70\% class, reflecting the expected weakening of collective flow in less central collisions.

In contrast, $T_\mathrm{kin}$ increases from central to peripheral collisions, rising from approximately 0.136~GeV at 10--20\% centrality to 0.172~GeV and 0.235~GeV in the 30--40\% and 60--70\% classes, respectively. This trend is consistent with a scenario in which particles in peripheral collisions decouple earlier at higher temperatures due to the shorter lifetime and smaller system size, whereas in more central collisions the system expands and cools for a longer period before freeze-out.

The event-by-event fluctuations $\sigma(\beta_{s})$ and $\sigma(T_\mathrm{kin})$ exhibit a pronounced increase from central to peripheral collisions. Specifically, $\sigma(\beta_s)$ rises from approximately 0.012 in 10--20\% centrality to 0.021 and 0.046 in the 30--40\% and 60--70\% classes, respectively, reflecting enhanced variability in the collective expansion strength in smaller systems. Similarly, $\sigma(T_\mathrm{kin})$ increases from around 0.002~GeV to 0.006~GeV and 0.020~GeV, indicating more pronounced thermal fluctuations at kinetic decoupling. This escalation of fluctuations in peripheral collisions can be attributed to the reduced system size and shorter lifetime.

\begin{figure*}
\centering
\includegraphics[width=0.95\textwidth]{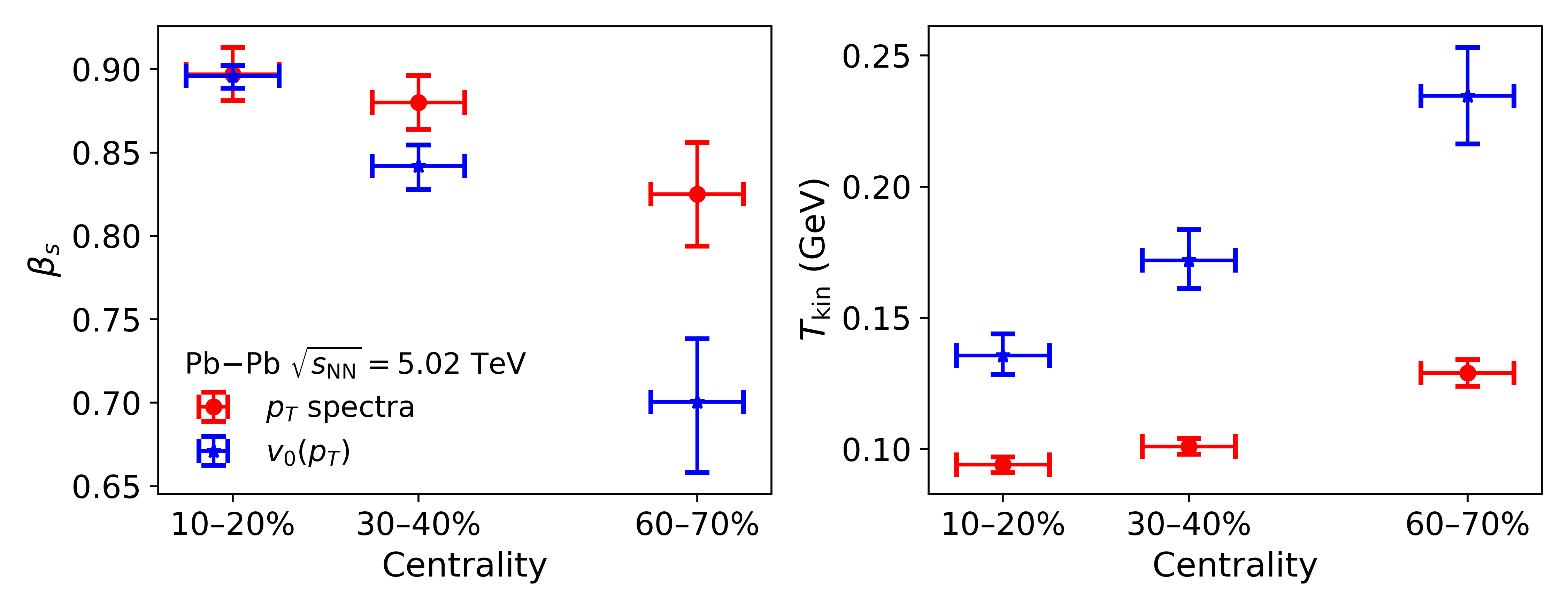}
\caption{Comparison of (left) $\beta_s$ and (right) $T_\mathrm{kin}$ values obtained in this analysis using $v_0(p_\mathrm{T})$ measurements with those extracted from blast-wave fits to $p_\mathrm{T}$ spectra reported in Ref.~\cite{ALICE:2019hno}. }
\label{fig:compare_pub}
\end{figure*}

Figure~\ref{fig:compare_pub} compares the values of $\beta_s$ and $T_\mathrm{kin}$ extracted in this analysis using $v_0(p_\mathrm{T})$ measurements with those obtained from blast-wave fits to $p_\mathrm{T}$ spectra in Ref.~\cite{ALICE:2019hno} (with $\beta_s$ values derived using Eq.~\ref{eq:betaS}). As shown in the left panel, the $\beta_s$ values from both methods are consistent within uncertainties, with deviations remaining below 2$\sigma$. In contrast, the right panel shows significant differences in the extracted $T_\mathrm{kin}$ values: the temperatures from the $v_0(p_\mathrm{T})$ analysis are systematically higher than those from the $p_\mathrm{T}$ spectra.

A key reason for this discrepancy lies in the differing sensitivity of the two observables to resonance decays. The $p_\mathrm{T}$ spectra include contributions from short-lived resonances that decay into lighter hadrons, particularly at low $p_\mathrm{T}$. These decay products tend to broaden and soften the spectra, leading to a lower apparent freeze-out temperature when fitted with a blast-wave model. In contrast, $v_0(p_\mathrm{T})$ is extracted from two-particle correlation functions using a pseudorapidity gap, which suppresses short-range correlations, including those from resonance decays. Since resonance decay pairs typically emerge close in angle and rapidity, their contribution to $v_0(p_\mathrm{T})$ is significantly reduced. As a result, the reduced influence of resonances in $v_0(p_\mathrm{T})$ may offer a different perspective and a more direct probe of the thermal conditions at kinetic freeze-out.

\section{Summary and Outlook}
\label{sec:summ}

We have investigated the transverse momentum dependence of the radial flow observable $v_0(p_\mathrm{T})$, which characterizes the isotropic expansion of the hot and dense medium created in high-energy heavy-ion collisions. Our analysis focuses on identified particles—pions, kaons, and protons—using a blast-wave model that incorporates Gaussian event-by-event fluctuations in the transverse expansion velocity and kinetic freeze-out temperature. 

The model qualitatively reproduces key features of the $v_0(p_\mathrm{T})$ distributions observed in recent ALICE data. Furthermore, using this model, we provide insight into how (a) radial flow, (b) radial flow fluctuations, (c) kinetic freeze-out temperature, and (d) temperature fluctuations affect the $p_\mathrm{T}$ dependence of the observable. For smaller mean values of $\beta_s$, $v_0(p_\mathrm{T})$ shows little mass dependence; however, increasing $\beta_s$ leads to a clear mass ordering, reflecting stronger radial flow. The $v_0(p_\mathrm{T})$ transitions from negative values at low $p_\mathrm{T}$ to positive values above a species-dependent threshold $p_\mathrm{T,sep}$, which shifts to higher $p_\mathrm{T}$ with increasing $\beta_s$. Fluctuations in $\beta_s$ and $T_\mathrm{kin}$ amplify the magnitude of $v_0(p_\mathrm{T})$, particularly at higher $p_\mathrm{T}$, with heavier particles exhibiting greater sensitivity.

Using Bayesian parameter estimation, we extracted blast-wave parameters and their event-by-event fluctuations by fitting $v_0(p_\mathrm{T})$ to ALICE measurements in Pb--Pb collisions at $\sqrt{s_\mathrm{NN}}=5.02$ TeV for three centrality intervals: 10--20\%, 30--40\%, and 60--70\%. The results reveal a systematic decrease in $\beta_s$ and increase in $T_\mathrm{kin}$ from central to peripheral collisions, consistent with reduced collective expansion and earlier freeze-out in smaller systems. This general anti-correlation between $\beta_s$ and $T_\mathrm{kin}$ has been observed in heavy-ion collisions when these parameters are extracted using hadron $p_\mathrm{T}$ spectra. The fluctuation widths $\sigma(\beta_s)$ and $\sigma(T_\mathrm{kin})$ also increase toward peripheral collisions, indicating enhanced event-by-event variability. 

Model predictions based on the best-fit parameters describe the experimental data well at low $p_\mathrm{T}$, with the strongest agreement observed for protons (over a larger $p_\mathrm{T}$ range), followed by kaons and pions. These results provide new insights into the nature of collective expansion and freeze-out dynamics, highlighting the role of flow strength, thermal motion, and their fluctuations in shaping $v_0(p_\mathrm{T})$ across different collision centralities.

An important extension of this work would be to apply $v_0(p_\mathrm{T})$ modeling to smaller systems such as pp and p--Pb collisions. The nature and origin of collectivity in these systems have been under active investigation since the first observation of long-range, near-side correlations in two-particle distributions~\cite{CMS:2010ifv, CMS:2012qk}. Although the blast-wave model can, in principle, be adapted to such systems by constraining the system size and freeze-out conditions, interpreting $v_0(p_\mathrm{T})$ in this context may require incorporating additional physics, such as initial-state momentum correlations or non-equilibrium dynamics, to account for the different underlying mechanisms compared to heavy-ion collisions. Future studies should also consider incorporating event-by-event geometry and system-size fluctuations, which are not included in the current blast-wave implementation but may significantly influence $v_0(p_\mathrm{T})$, especially in small collision systems.

Moreover, at lower collision energies—such as those explored in the RHIC Beam Energy Scan program—the changing balance between thermal pressure and baryon density, along with possible critical fluctuations near the QCD phase transition, may imprint distinct signatures on $v_0(p_\mathrm{T})$. Studying $v_0(p_\mathrm{T})$ systematically across different systems and energies could thus offer new insights into the interplay between radial flow, fluctuations, and the underlying properties of QCD matter.

\begin{acknowledgments}
This work is supported by the Department of Atomic Energy (DAE), Government of India. We also acknowledge the use of Garuda HPC facility and Kanaad facility at the School of Physical Sciences, NISER. We are grateful to Jean-Yves Ollitrault, Björn Schenke, Chun Shen, Derek Teaney, Tribhuban Parida, and Rupam Samanta for insightful discussions on the new observable $v_0(p_\mathrm{T})$. We thank the ALICE Collaboration for making the $v_0(p_\mathrm{T})$ data publicly available. The Bayesian inference was performed using Python-based MCMC routines.

\end{acknowledgments}

\bibliographystyle{utphys}
\bibliography{references}

\end{document}